\newif\ifsubmode
\submodefalse
 
\newif\ifprintfig
\printfigfalse
 
\ifsubmode
  \documentclass{aastex}
  \received{}
  \accepted{}
  \journalid{}{}
  \articleid{}{}
\else
  \documentclass[preprint]{aastex}
  \usepackage{epsfig}
\fi
 
\shortauthors{B\"oker et al.}
\shorttitle{Bulges in Late Hubble Types}

\def\spose#1{\hbox to 0pt{#1\hss}}
\def\lta{\mathrel{\spose{\lower 3pt\hbox{$\sim$}}
    \raise 2.0pt\hbox{$<$}}}
\def\gta{\mathrel{\spose{\lower 3pt\hbox{$\sim$}}
    \raise 2.0pt\hbox{$>$}}}
\newcommand{\ea}{et al.}

\newcommand{\s}{\>{\rm s}}
\newcommand{\kms}{\>{\rm km}\,{\rm s}^{-1}}
\newcommand{\pc}{\>{\rm pc}}
\newcommand{\kpc}{\>{\rm kpc}}

\newcommand{\lsun}{\>{\rm L_{\odot}}}

\newcommand{\as}{ ^{\prime\prime}}
 
\newcommand{\bdm}{\begin{displaymath}} 
\newcommand{\edm}{\end{displaymath}}
\newcommand{\beq}{\begin{equation}} 
\newcommand{\eeq}{\end{equation}} 
\newcommand{\bit}{\begin{itemize}} 
\newcommand{\eit}{\end{itemize}} 
\newcommand{\ben}{\begin{enumerate}} 
\newcommand{\een}{\end{enumerate}}
\newcommand{\bfi}{\begin{figure}[htb]} 
\newcommand{\bpfi}{\begin{figure}[p]}

\begin{document}

\title{Searching for Bulges at the End of the Hubble 
Sequence\altaffilmark{1}}

\author{Torsten B\"oker\altaffilmark{2}} 
\affil{Space Telescope Science Institute, 3700 San Martin Drive, 
       Baltimore, MD 21218}
\email{boeker@stsci.edu}

\author{Rebecca Stanek}
\affil{University of Michigan, Department of Astronomy, Dennison Building,
       Ann Arbor, MI 48109}
\email{rstanek@astro.lsa.umich.edu}

\author{Roeland P.~van der Marel} 
\affil{Space Telescope Science Institute, 3700 San Martin Drive, 
       Baltimore, MD 21218}
\email{marel@stsci.edu}


\altaffiltext{1}{Based in part on observations made with the NASA/ESA 
{\it Hubble Space Telescope}, obtained at the Space Telescope Science 
Institute, which is operated by the Association of Universities for 
Research in Astronomy, Inc., under NASA contract NAS 5-26555. These
observations are associated with proposal \#\,8599.}

\altaffiltext{2}{On assignment from the Space Telescope Division of the
    European Space Agency (ESA).}


 
\ifsubmode\else
\newpage\fi
 
 
\ifsubmode\else
\baselineskip=14pt
\fi


\begin{abstract} 
We investigate the stellar disk properties of a sample of 19
nearby spiral galaxies with low inclination and late Hubble type (Scd
or later).  We combine our high-resolution {\it Hubble Space
Telescope} I-band observations with existing ground-based optical
images to obtain surface brightness profiles that cover a high dynamic
range of galactic radius. Most of these galaxies contain a nuclear
star cluster, as discussed in a separate paper.  The main goal of the
present work is to constrain the properties of stellar bulges at these
extremely late Hubble types.  We find that the surface brightness
profiles of the latest-type spirals are complex, with a wide range in
shapes.  We have sorted our sample in a sequence, starting with
``pure'' disk galaxies (approximately 30\% of the sample).  These
galaxies have exponential stellar disks that extend inwards to within
a few tens of pc from the nucleus, where the light from the nuclear
cluster starts to dominate. They appear to be truly bulge-less
systems. Progressing along the sequence, the galaxies show
increasingly prominent deviations from a simple exponential disk model
on kpc scales. Traditionally, such deviations have prompted
``bulge-disk'' decompositions. Indeed, the surface brightness profiles
of these galaxies are generally well fit by adding a second
(exponential) bulge component. However, we find that most surface
brightness profiles can be fit equally well (or better) with a single
S\'ersic-type $R^{1/n}$ profile over the entire radial range of the
galaxy, without requiring a separate ``bulge'' component. We warn in 
a general sense against identification of bulges solely on the basis 
of single-band surface brightness profiles.
Despite the narrow range of Hubble types in our sample, the surface
brightness profiles are far from uniform. The differences between the
various galaxies appear unrelated to their Hubble-types, thus
questioning the usefulness of the Hubble-sequence for the
sub-categorization of the latest-type spirals.
A number of galaxies show central
excess emission on spatial scales of a few hundred parsec that can be
attributed neither to the nuclear cluster, nor to the S\'ersic-type
description of the stellar disk, nor to what one would generally
consider to be a bulge component. The origin of this light component
remains unclear. 
\end{abstract}




\section{Introduction}\label{intro}

The question whether the morphology of galaxies is imprinted by the
initial conditions of their formation or rather determined by secular
evolution remains a subject of intense debate. The existence of the
Hubble sequence has for many years provided important constraints on
this issue. In very simple terms the Hubble sequence tells us that
galaxies are made up of two components: a bulge and a disk. The
canonical view of these components has long been that bulges have
$R^{1/4}$ surface brightness profiles \citep{dev48} while disks have
exponential surface brightness profiles. As one goes from early-type
to late-type galaxies one goes from galaxies that are bulge-dominated
to galaxies that are disk-dominated. While this simplistic
interpretation of the Hubble sequence has definite value, reality is
considerably more complicated.

In recent years, our views of the Hubble sequence have evolved and
gained more nuance. For elliptical galaxies it has become clear that
they are not necessarily pure bulge systems: many elliptical galaxies
contain embedded disks.  There is evidence from other information
(e.g., kinematics) that elliptical galaxies form a heterogeneous class
of galaxies that may have formed in different ways
\citep[e.g.][]{kor96}. For spiral galaxies a clearer understanding has
developed of their bulge properties. High-resolution imaging - both
from the ground \citep[e.g.][]{dej96a}, and with the {\it Hubble Space
Telescope} \citep[HST,][]{car98} - has shown that the central surface
brightness profile (SBP) of many late-type spirals cannot be fit by
the classical $R^{1/4}$ law that is well suited to describe the bulge
profiles of early-type spirals. Instead, the SBPs of
many late-type spirals rise above the extrapolation of the exponential
disk in a way that can be well described by a second exponential
\citep{and94}. This has led to the now popular view that spiral bulges 
come in two flavors: on the one hand, the classical $R^{1/4}$ bulges
which are mostly observed in early-type spirals, and on the other the
``pseudo-bulges'' \citep{kor93} or ``exponential bulges''
\citep[][]{car98} which are prevalent in later Hubble types. In reality 
there is probably a continuum of properties, instead of a dichotomy.
When $R^{1/n}$ profiles \citep{ser68} are fit to available SBPs, the
profile shape parameter spans the full range of values $1 \leq n \leq
6$; the profile shape parameter correlates with both Hubble type and
bulge-to-disk ratio of the galaxy\footnote{Obviously, Hubble type and
bulge-to-disk ratio are themselves strongly correlated through the
definition of the Hubble sequence.}, in the sense that spiral galaxies
with earlier Hubble type have bulges with higher $n$ values
\citep*{and95,gra01}.

The existence of different types of bulges in disk galaxies can be
plausibly explained in the context of popular scenarios for the
formation and secular evolution of galaxies. The classical massive
$R^{1/4}$ law bulges fit in with the ``primordial collapse'' formation
scenario first suggested by \cite*{egg62}, in which the bulge forms
during the initial collapse of a galaxy-sized density perturbation, and
later ``acquires'' a disk through accretion processes. By contrast, the
pseudo-bulges may have formed by secular evolution of a pre-existing
disk, so that they formed after the disk, out of disk material. 
Some support for this scenario comes from the fact that pseudo-bulges 
are dynamically similar to their host disks \citep{kor93}. Plausible 
secular evolution scenarios include the accretion of satellite galaxies 
\citep{agu01}, buckling instabilities in a stellar bar \citep*{rah91}, 
and the disruption of a stellar bar through the accumulation of a central 
mass concentration \citep*{nor96}. Many discussions of these and 
related topics can be found in the review by \cite*{wys97} and in the 
proceedings of the recent workshop on `The Formation of Galactic Bulges' 
\citep*{car99}.

In the present paper we study the presence and properties of bulges in
the very latest-type spiral galaxies (Scd or later). This is an
important topic for several reasons. First, these galaxies are
generally classified as very late type spirals because they do not
have a very prominent bulge. As a result, many observational studies
of bulges have avoided these galaxies. Second, it has become clear
from recent work with HST that the majority of spiral galaxies contain
a central star cluster. In the very latest-type spiral galaxies we
find that $\sim 75$\% of the galaxies contain such a star cluster
\citep[][hereafter paper I]{boe02}. In late Hubble types, these clusters 
are easily mistaken for a small bulge when observed from the ground,
even in good seeing conditions. So the bright, compact ``bulges'' in
late-type spirals which were used as a classification criterion in the
original work of \cite{hub26,hub36} may in fact be dense star clusters
occupying the photocenter of the galaxy. The purpose of this paper is
to shed some light on these issues. In particular, we investigate
whether the very latest-type spirals are completely bulgeless, whether
they show excess light above the constant scale-length disk, and if
so, whether this in fact implies the presence of a separate entity
which could rightfully be called a bulge.

HST resolution is needed to separate the luminous nuclear star cluster
from a putative bulge. Our I-band snapshot survey of late-type
spiral galaxies conducted with the {\it Wide Field and Planetary
Camera 2} (WFPC2) and discussed in Paper~I therefore forms the basis of our
analysis. We complement the HST observations with ground-based data that
extends to larger radii. The paper is organized as follows: in
\S~\ref{sec:data}, we describe the data and the analysis methods that
form the basis of our work.  The results of our analysis are
summarized in \S~\ref{sec:results}.  We discuss the implications of
our findings, and present our conclusions in
\S~\ref{sec:disc}.
 
\section{Data} \label{sec:data}

\subsection{HST Data} \label{sec:hst}
The target list and selection criteria for the HST sample are 
described in paper I. In brief, we selected 113 nearby 
($\rm v_{\rm helio} < 2000~km/s$) galaxies with low inclination
($\rm R_{25} \equiv log(a/b) < 0.2$) and Hubble type between Scd 
and Sm ($\rm 6 < T < 9$). 77 galaxies have been observed to date
with WFPC2 on board HST in the F814W filter which is similar to 
Johnson I-band. In all cases, the integration time was $600\,s$.
After standard data reduction to correct for instrumental effects
and cosmic rays, we performed an isophotal analysis using the 
{\tt ellipse} task in IRAF. For all galaxies studied in the present
analysis, the ellipse center, ellipticity, and position angle 
were allowed to vary freely from isophote to isophote. 

The WFPC2 SBPs have been published in Figures~1 and 2 of paper I. 
In this study, we discuss the 19 galaxies for which we
were able to find ground-based, wide-field surface brightness data
for comparison. Our focus lies on the structural properties of the 
disks rather than the photometry of the nuclear clusters (which was
the subject of paper I).

\subsection{Ground-Based Surface Photometry}\label{sec:comp}

Due to the short integration time ($600\s$) of the WFPC2 snapshot
observations, the signal-to-noise ratio is generally insufficient to
make use of the WF chips in our analysis. It is therefore prudent to
ask whether the rather limited field-of-view of the PC chip yields a
representative view of the galaxy disk.  In order to address this
question, we searched the literature for published wide-field
observations of our sample galaxies. In general, these are scarce,
most likely due to the less than spectacular star formation activity
and rather low surface brightness of the latest Hubble type spirals.
In this context, it is worth pointing out that all-sky surveys such as
2MASS do not reach the required sensitivity to map the outer disk of
most of our sample galaxies, and hence are not useful for our
purposes.

We found published wide-field SBPs for 19 of our
sample galaxies which are listed in Table~\ref{tab:sample}. 
All but one (NGC\,4904) of these 19 galaxies
harbor a nuclear cluster as described in paper I. The wide-field 
surface brightness data for 18 galaxies are based on images in the 
{\it Photometric Atlas of Northern Bright Galaxies} \citep[PANBG, ][]{panbg}.
The digitized profiles, which have been studied in detail by
\cite*{bag98}, were kindly provided to us in machine readable form 
by M. Hamabe. The PANBG images were taken in the V-band, with SBPs
derived from cuts along the major and minor semi-axes of the galaxy,
typically extending out to radii between $100\as$ and $250\as$.  For
comparison to the high-resolution WFPC2 data, we use (in most cases)
the average of the two semi-major axis cuts. For some galaxies, visual
inspection of the images and profiles in the PANBG showed that one or
both semi-major axis cuts appeared affected by large-amplitude,
small-scale variations. In these cases, we either used the remaining
semi-major axis cut, or the average of the {\it semi-minor} axes,
scaled by the ellipticity of the envelope ellipse (the elliptical fit
to the isophote contour at $\mu_{\rm V} = 25$) listed in column\,13 of
the PANBG.  The remaining galaxy (NGC\,2805) was part of the sample of
\cite{ken84}, who used a red passband centered at 6500\AA. 

Figure~\ref{fig:baghst} shows the available data for the 19 galaxies.
For each galaxy, the left panel contains the WFPC2 I-band SBPs; 
the right panel contains the PANBG data for each galaxy. The
ordering of the galaxies in the figure and the solid lines in the
panels are discussed in \S~\ref{subsec:seq} below. In order to
normalize both datasets, we shifted the (V-band) PANBG profiles by the
amount needed to match the average I-band surface brightness in the
region between $8\as$ and $12\as$. Typical offsets were in the range
$1.0 \leq V-I \leq 1.5$. This is not unreasonable, given that the
typical integrated color of galaxies of these Hubble types is $V-I =
1.0 \pm 0.2$ \citep[e.g.][]{dej96b}. The colors that we infer appear
to be somewhat too red, probably indicating a systematic offset
between the two datasets in their photometric calibration. However, in
this study, we are interested mainly in the structural properties of
the light distribution and this problem therefore does not affect our
analysis.

There are a few limitations to the combined data set thus obtained.
For one, the PANBG profiles were derived from simple cuts along the
galaxy major or minor axis, rather than from azimuthally averaged
isophote fitting methods (which is what we used for the HST data). The
PANBG profiles are therefore affected by spiral arms, dust lanes, star
forming regions, or other asymmetries, even though we made every
attempt to minimize these by selecting the least-affected
semi-axes. Also, possible mass-to-light variations or color gradients
could complicate the direct comparison of the two SBPs. However,
the results of \cite{dej96b} and \cite{mat97} have shown that
color gradients in late-type disks are generally small, with
variations in (B-I) of $\lta 0.2\>$mag over a few scalelengths. It 
thus appears that - except for the very center - the stellar 
populations of late-type disks are rather uniform, and our 
approach of profile matching seems justified. 
In summary, the PANBG profiles should yield a fairly robust impression of the
SBP of the outer disk. They are certainly useful for comparison to our
WFPC2 images, especially to test whether the central $15\as$ are
representative for the profile of the stellar disk.

\section{Results \label{sec:results}}

\subsection{Surface Brightness Profile Shapes}\label{subsec:seq}

Although the galaxies in our sample were selected as a fairly
homogeneous group with respect to Hubble type, distance, and
inclination, the SBPs show a wide range of shapes. To
address this quantitatively, we started by fitting for each galaxy an
exponential profile to the I-band SBPs inside the PC field. We will
refer to this as the `inner exponential fit'. In this fit we always
excluded the nuclear cluster, as well as any `central excess emission'
(the definition and nature of which is discussed below), if
present. The fits are shown as solid lines in the left panel of each
plot. We show the same exponential fit also in the right panel of each
plot (so the solid line in the right panel is {\it not} a fit to the
PANBG data, but only the continuation of the fit in the left panel).

The extent to which the inner exponential fit describes the data at
the different radii differs considerably for the different
galaxies. To illustrate this point, we have ordered the galaxies in
Figure~\ref{fig:baghst} in a rough sequence, according to the
following scheme. The sequence starts with galaxies for which there is
no evidence for a bulge of other inner component in the SBP, except
for the nuclear star cluster (which has a typical HWHM radius of
$5-10\pc$; paper I). This group itself is somewhat heterogeneous. For
some galaxies the inner exponential fit yields a good description of
the large-scale stellar disk as measured by the PANBG profiles (e.g.,
NGC\,428). These galaxies are fairly well described by a single
exponential. For other galaxies (e.g., NGC\,4299), the outer disk
falls below the inner exponential fit, which implies that the SBP at
small radii falls {\it below} the inward extrapolation of the outer
disk. This is exactly opposite to what one would expect if a bulge
were present (see~\S~\ref{subsec:really} below). We interpret both
types of galaxy as pure disk systems. Interestingly, these pure disks
have SBPs that can differ from a single exponential.

Progressing along the sequence, the profiles show an increasing amount
of light in excess of the inner exponential fit on radial scales of a
few hundred pc (the affected radii were excluded from the inner
exponential fit). We refer to this light as `central excess
emission'. This emission has a distinct profile shape: while the
overall galaxy profile steepens towards the center, the excess
component flattens towards the center (see, e.g., NGC\,275, NGC\,2139,
or NGC\,3346). This is different from what one would expect for a
bulge.  Bulges tend to have $R^{1/n}$ profiles with $n \gta 1$
\citep*{and95,gra01}. Such profiles steepen towards the
center, when plotted as a function of (linear) $r$. Consequently, we
interpret the central excess emission as a phenomenon that is
different from a central bulge. We discuss this component further in
\S~\ref{subsec:exc}.

Approximately halfway along the sequence the outward extrapolation of
the inner exponential fit begins to underpredict considerably the true
brightness of the outer disk (e.g., NGC\,5068). Conversely, this
implies that there is excess light in the central few kpc over the
inward extrapolation of the outer disk. This is the component that is
traditionally called a bulge. We discuss the nature of this component
further in \S~\ref{subsec:really}.
 
Given the narrow range of Hubble types of our sample (between Scd and
Sm), and the uniformity in inclination (less than $40\deg$ from
face-on), the variety of profile shapes is somewhat surprising. In
addition, there seems to be no clear trend of position along our
sequence with either Hubble type or galaxy luminosity. For example,
the profiles of NGC\,275 and NGC\,1042 are very different, even though
both galaxies have nearly identical Hubble type (SB(rs)cd
vs. SAB(rs)cd) and total blue magnitude ($\rm M_B = -18.8$). Clearly,
late-type spirals are not simple systems, their morphologies (and
possibly evolutionary states) can differ drastically.

\subsection{Disks and Bulges}\label{subsec:really}
A considerable source of confusion in discussions of bulges is that
different definitions are often used when referring to a bulge. 
%
Modern theorists tend to think of a bulge as a
kinematically hot component with an extended three-dimensional
structure. However, the observational definition of a bulge is often a
different one, and relies on the assumption that bulges 
have different SBPs than disks. The standard wisdom is
that a disk can be well fit by an exponential profile, $I(r)\propto
e^{-r/r_d}$. By contrast, bulges can be described empirically with
either a second exponential with different scale length
\citep[e.g.][]{dej96a}, a de Vaucouleurs $R^{1/4}$ profile
\citep[e.g.][]{bor81,bag98}, or most generally, a S\'ersic $R^{1/n}$
profile \citep[e.g.][]{gra01}. 

%
%

If one defines a bulge as `the central concentration of 
mass [or light] in excess of the inward extrapolation of the outer, 
constant scale-length, exponential disk' \citep*{car99}, one must make 
sure that exponential profiles indeed provide a highly accurate
description of galaxy disks. Here one is on shaky ground. The exponential
model has been widely used to describe galaxy disks because it
provides a fairly good fit to many observed galaxy profiles
\citep[e.g.][]{fre70,ken84,dej96a}. However, the fits are generally
restricted to the region outside the central few kpc, so that it is
actually very hard to know whether disks remain exponential all the way
into the center. The disks of many late-type spirals in fact show
sharp, well-defined ``breaks'' in their brightness profiles, which
separate sections of the disk that follow exponential profiles with
different scale lengths \citep{bos93,mat97}. So observationally, it is
not at all clear how photometry alone can distinguish between an
exponential bulge and a somewhat steeper disk section in the inner
part of the galaxy. 

Theoretical considerations also do not clinch the
argument. There is no theory that unequivocally predicts that disks
must be exponential. As an `a posteriori' explanation,
(semi-)analytical models that invoke angular momentum redistribution
via viscosity-driven radial gas flows can provide a plausible
mechanism for building profile shapes close (but not identical) to an
exponential. This has been demonstrated in the initial work by
\cite{lin87} and \cite{yos89}, and recently confirmed by the refined
study of \cite{fer01}. An alternative scenario which invokes
self-propagating stochastic star formation in a disk of constant
density atomic hydrogen gas \citep*{sei84} also results in brightness
profiles that approximate the exponential shape. However, all these
models are dependent on a number of parameters, which can be
fine-tuned to reproduce a range of profile shapes. In fact, most model
profiles presented by \cite{yos89} show significant
curvature at all radii. It is therefore not obvious that a simple
exponential is a good analytic model for disks to begin with.

In order to demonstrate the ambiguity of profile
decompositions, and the ensuing interpretation of the derived
``bulge'' properties, we have compared two different fits to the
combined (WFPC2 plus PANBG) datasets, excluding again the nuclear
cluster. The two models are: (a) a single S\'ersic profile; and (b)
the sum of two exponentials of different scale lengths. Interestingly,
we find that most galaxies in the second half of our sequence are
equally well fit with either model, thus questioning the need for a
``bulge'' component. This is demonstrated in Figure~\ref{fig:comps}
which compares these two fits for five galaxies with apparent
disk-bulge transitions. In most cases the two profiles provide equally
adequate fits, and for one galaxy, NGC\,5668, the S\'ersic profile
provides an obviously better fit. Only for one galaxy, NGC\,1042, does
a sum of two exponentials provide the better fit.

Overall, we find that roughly half of our sample galaxies show excess
light (over the inward extrapolation of the outer exponential disk)
that can be fit by a second exponential. This is similar to the
`pseudo bulges' or `exponential bulges' that have been reported by
other authors, mostly in earlier type spirals. However, the extent to
which this excess light is associated with a bona-fide bulge component
remains an open question. The SBPs of these galaxies are generally
well fit with S\'ersic profiles with $n \approx 1$--$2.5$ (as listed
in the panels of Figure~\ref{fig:comps}). In the absence of information
on kinematics or three-dimensional structure it is difficult to rule
out that we are dealing merely with non-exponential disks, or possibly
small-scale stellar bars which are known to also have exponential-like
SBPs \citep[e.g.][]{elm85}.

\subsection{The central light excess}\label{subsec:exc}

As described in \S~\ref{subsec:seq}, several galaxies of our sample
show evidence for central excess emission on scales of a few hundred
pc which does not show the steepening characteristic of bulges. In
order to investigate the nature of this emission, we have measured the
amount of excess light for four example cases (NGC\,275, NGC\,2139,
NGC\,3346 and NGC\,5584). We first determined by eye the radius
$r_{ap}$ at which the SBP starts to deviate from the inward
extrapolation of the exponential fit. We then analyzed the WFPC2
I-band images presented in paper~I with the {\it IRAF} photometry
package {\it apphot} in order to measure the total flux within a
circular aperture of radius $r_{ap}$, centered on the nuclear star
cluster. To give a better visual impression of the four case studies,
Figure~\ref{fig:four} reproduces their images presented in paper
I. Here, we have overlaid circles with radius $r_{ap}$ to indicate the
aperture over which we measure the excess emission. Also, we have
slightly modified the grey scale stretch to emphasize the disk
structure rather than the nuclear cluster.

From the total flux inside $r_{ap}$, we subtract the contribution of
the galaxy disk which is calculated by integrating the exponential fit
(the solid lines in Figure~\ref{fig:baghst}) inside $r_{ap}$.  We also
subtract the light from the nuclear cluster as listed in paper~I,
which usually constitutes only a small correction.
Table~\ref{tab:excess} summarizes the results of this analysis.  For
the four galaxies of Figure~\ref{fig:four}, the apparent magnitude of
the excess emission is in the range $15.2 \leq m_I \leq 17.4$ which
corresponds to luminosities between $2\cdot 10^7\lsun$ and $2\cdot
10^8\lsun$. This is between 10 and 100 times brighter than the median
luminosity of the nuclear star cluster in our sample ($\rm M_I =
-11.5$ or $1.6\cdot 10^6\lsun$, paper I), but constitutes only
between 1\% and 5\% of the total galaxy luminosity. It is
interesting that with absolute luminosities of $-16.7 \leq M_I \leq -14.2$ 
(using the distances in Table~\ref{tab:excess}), this excess emission 
appears to be a faint end continuation of the distribution of spiral 
bulge luminosities shown in Figure\,14 of \cite{gra01}, if one assumes 
a mean color of $\rm B-I \approx 1$.

The images in Figure~\ref{fig:four} reveal a variety of morphologies
for the excess emission. For NGC\,275, NGC\,5584, and NGC\,3346, the 
emission appears fairly smooth and follows the larger scale disk 
structure. However, this
is clearly not the case for NGC\,2139, for which the observed excess
emission (which is also by far the brightest in our sample) is due to
a prominent, bright star forming region which is elongated in the E-W
direction. Inspection of the WF chips shows that this structure is the
inward continuation of at least one spiral arm (the other direction is
off the WFPC2 field of view).  This underlines the danger of relying
solely on one-dimensional surface brightness data for evaluating
galaxy morphology.

\section{Summarizing Discussion \label{sec:disc}}

We have presented an investigation into the structural properties of
19 spiral galaxies with Hubble type between Scd and Sm. From a
combination of our high-resolution HST data and wide-field
ground-based images, we obtain surface brightness profiles (SBPs) for
the sample galaxies that cover a large dynamic range in galactic
radius. We use these profiles to study quantitatively the presence and
properties of bulges in spiral galaxies of the very latest Hubble
types. Of course, by the very definition of the Hubble sequence, one
does not expect very prominent bulges in these galaxies. However, it
has been somewhat of an open question whether bulges are present at
all in these galaxies. It has been realized only recently that
space-based resolution is required to properly address this question.
The nuclear morphologies of most spiral galaxies are complex, and a
large fraction of spiral galaxies has a nuclear star cluster. These
clusters are easily mistaken for ``compact bulges'' when observed with
ground-based resolution. Previous studies with the HST have focused in
majority on earlier Hubble types, and the present study is the first
to focus exclusively on the very latest Hubble types.

Approximately 30\% of the sample galaxies seem to be more-or-less
``pure'' exponential disks, without any type of a central bulge. Our
sample was selected to focus on face-on galaxies, but studies of
edge-on galaxies support the view that disk galaxies can indeed be
completely bulgeless. \cite*{mat99} and \cite*{mat00} have studied
edge-on ``super-thin'' galaxies such as UGC\,7321, and have
demonstrated that in these galaxies there is no evidence for a
spheroidal component.

Despite being disk-dominated systems, most galaxies in our sample 
have SBPs that cannot be well fit by a
single exponential, in the sense that the surface brightness in the
central few kpc exceeds the inward extrapolation of the outer
exponential disk. This has generally been found for other samples of
spirals as well, and has generally prompted the addition of a bulge
component to analytic models of the SBP. In particular, numerous
studies have shown that the SBPs of intermediate- to late-type spirals
can be well fit by a sum of two exponentials. The inner exponential
has traditionally been interpreted as an `exponential' bulge or
`pseudo' bulge. Such bulges can be (qualitatively) explained
theoretically as a result of secular evolution of the disk. While this
may be correct, we point out that this is not an unambiguous
interpretation. In the absence of information on three-dimensional
structure or dynamics there is no guarantee that one is dealing with a
bona-fide bulge component. We explicitly illustrate this point for the
galaxies in our sample that are not well fit by a single exponential.
Indeed, the SBPs of these galaxies are generally well fit by a sum of
two exponentials. However, most profiles can be described at least
equally well with a S\'ersic-type $R^{1/n}$ model over the entire
radial range (outside the nuclear star cluster). The shape parameter
$n$ is in the range $1 \lta n \lta 2.5$, which is not unrealistically
large. As we have discussed in \S\ref{subsec:really}, there is no a priori 
theoretical reason for the SBPs of disk
galaxies to be pure exponentials; S\'ersic-type profiles are in some sense
equally arbitrary as the model of choice. So it may well be that we are simply 
dealing with non-exponential disks in most of our sample galaxies, and 
that bona-fide bulges are rare at these Hubble types.

We have found that a number of late-type galaxies show central excess
emission on spatial scales of a few hundred parsec that can be
attributed neither to the nuclear cluster, nor to the S\'ersic-type
description of the stellar disk, nor to what one would generally
consider to be a bulge component. The origin of this light component
remains unclear.

One of the interesting findings from our work is that, despite the
narrow range in Hubble type, the SBPs of the sample galaxies are far
from uniform. Our study finds no systematic trends in the structural
properties with morphological type. The exact Hubble type of spirals
between Scd and Sm appears somewhat arbitrary, in the sense that it
provides little information about the presence and relative importance
of galaxy bulges. It is quite possible that this may be due to the
presence of nuclear clusters, which may have played an important role
in the morphological classification in photographic catalogs. In
general our results fit in well with the picture that emerged from a
ground-based imaging study of 49 late-type spirals by
\cite{mat97}. They found that late-type spirals
exhibit a diverse array of structural properties and morphologies,
even in galaxies with otherwise similar parameters, and they concluded
that bulges are often very weak or non-existent. In studies of this
kind one does need to be concerned about selection effects and
observational bias. The \cite{mat97} sample was selected for low
galaxy luminosity ($M_V \gta -18.8$), and one might worry that it did
not provide a representative view of the family of very late-type
spirals. Our galaxy sample, on the other hand, has been selected only
for Hubble type, distance, and inclination. It thus includes more
luminous (and presumably more massive) galaxies than the
\cite{mat97} sample. The blue absolute luminosity of the galaxies 
in our sample ranges from $M_B = -17.4$ (NGC\,2552, NGC\,4701) to $M_B
= -20.5$ (NGC\,2805). However, we find no systematic correlation
between absolute luminosity and position along the sequence of SBP
shapes discussed in \S\ref{subsec:seq}. This suggests that more
luminous late-type spirals are not systematically different from their
faint-end relatives, and that presumably the \cite{mat97} results are
valid for most late-type spirals.

Late type spiral galaxies are in most ways ``normal'' spiral galaxies.
Their angular momenta and rotation velocities are not atypical, as
demonstrated by a recent survey of optical rotation curves
\citep{mat02}. They are in some sense the dynamically simplest type 
of disk galaxies. They are (mostly) disk-dominated and often have only
faint spiral arm structure or even no detectable density perturbations
at all. Yet, we are clearly a long way from understanding their
formation and evolution. The latest-type disk galaxies provide the
most stringent observational constraints on the well-known angular
momentum problem in cold dark matter (CDM) galaxy formation models
\citep[e.g.][]{nav00}. Because of their unevolved disks and apparent
history of only modest star formation, they are a challenge for
proposed solutions for the angular momentum problem which rely on
energy feedback from supernova explosions \citep{sil01}. To gain a
better understanding of these issues it will be important to continue
to improve our knowledge of the structure of late-type spiral
galaxies. Observational studies such as the one presented here will
continue to be essential.

\acknowledgements

We are grateful to M. Hamabe for providing us with the digital surface
brightness data from the PANBG, to Roelof de Jong for helpful
discussions, and to the anonymous referee for
useful comments.  R. S. acknowledges a grant from the
STScI summer student program.  Support for proposal \#\,8599 was
provided by NASA through a grant from the Space Telescope Science
Institute, which is operated by the Association for Research in
Astronomy, Inc., under NASA contract no. NAS 5-26555.
This research has made use of the NASA/IPAC Extragalactic Database (NED) 
which is operated by the Jet Propulsion Laboratory, California Institute of 
Technology, under contract with NASA. 
It has also benefited greatly from use of the Lyon-Meudon
Extragalactic Database (LEDA, http://leda.univ-lyon1.fr).
%


\newpage
\figcaption[hst_vs_bag1.ps]{\label{fig:baghst}
  Left: HST/WFPC2 I-band surface brightness profile (circles), and the
  best-fitting exponential (line). For all galaxies, the datapoints 
  cover the field of view of the PC chip, i.e. about $15\as$ in radius. 
  Right: ground-based surface
  brightness profile (PANBG), corrected to the I-band as described in
  the text. The line in the right panel shows the same exponential
  profile as in the left panel; it is not a fit to the ground-based
  data. 
   }
\figcaption[2805_comp.ps]{\label{fig:comps}
  Comparison of model fits to the combined HST and ground-based 
  surface brightness data for five sample galaxies. For each object,
  the top panel shows the best-fiting sum (solid line) of two exponentials
  with different scale lengths (dashed lines). The bottom panel shows the
  best-fitting S\'ersic -type profile; its shape parameter $n$ is given
  in the top right corner. 
   }
\figcaption[four_mbgals.ps]{\label{fig:four}
  The WFPC2 I-band images of four galaxies with a central light excess
  as described in \S~\ref{subsec:seq}. Shown
  is the PC chip only, with a field-of-view of about $34\as$. The 
  North-East orientation is indicated by the arrow in the top right 
  corner of each image, with the arrow tip pointing north. The line
  in the top left corner indicates a scale of $1\kpc$. The circles
  denote the aperture within which the SBP
  shows excess emission above the exponential disk 
  (see Figure~\ref{fig:baghst}).
   }

\newpage
\begin{deluxetable}{lcccccc}
\tablecaption{The Sample \label{tab:sample}}
\tablewidth{0pt}
\tablehead{
\colhead{(1)} & \colhead{(2)} & \colhead{(3)} & \colhead{(4)} &
\colhead{(5)} & \colhead{(6)} & \colhead{(7)}   \\
\colhead{Galaxy} & \colhead{R.A.}  & \colhead{Dec.} & \colhead{$\rm v_z$} & 
\colhead{Type} & \colhead{$\rm m_B$} & \colhead{$\rm d_{MA}$}  \\
 & \colhead{(J2000)} & \colhead{(J2000)} & \colhead{[km/s]} &  & \colhead{[mag]} & 
 \colhead{[arcmin]}     
}
\startdata
NGC\,275    & 00 51 04.20 & -07 04 00.0 & 1681 & SB(rs)cd pec & 13.16 & 1.5   \\
NGC\,428    & 01 12 55.60 & -00 58 54.4 & 1130 & SAB(s)m   & 11.91 &  4.1  \\
NGC\,1042   & 02 40 23.63 & -08 25 59.8 & 1271 & SAB(rs)cd & 12.50 &  4.7  \\
NGC\,2139   & 06 01 07.90 & -23 40 21.3 & 1649 & SAB(rs)cd & 11.99 &  2.6  \\
NGC\,2552   & 08 19 20.14 & +50 00 25.2 &  695 & SA(s)m?   & 12.56 &  3.5  \\
NGC\,2805   & 09 20 24.56 & +64 05 55.2 & 1968 & SAB(rs)d  & 11.52 &  6.3   \\
NGC\,3346   & 10 43 38.90 & +14 52 18.0 & 1315 & SB(rs)cd  & 12.41 &  2.9  \\
NGC\,3423   & 10 51 14.30 & +05 50 24.0 & 1025 & SA(s)cd   & 11.59 &  3.8  \\
NGC\,3445   & 10 54 35.87 & +56 59 24.4 & 2245 & SAB(s)m   & 12.90 &  1.6  \\
NGC\,4027   & 11 59 30.50 & -19 15 44.0 & 1588 & SB(s)dm   & 11.66 &  3.2  \\
NGC\,4299   & 12 21 40.90 & +11 30 03.0 &  306 & SAB(s)dm: & 12.88 &  1.7   \\ 
NGC\,4540   & 12 34 50.90 & +15 33 06.9 & 1383 & SAB(rs)cd & 12.44 &  1.9  \\
NGC\,4701   & 12 49 11.71 & +03 23 21.8 &  768 & SA(s)cd   & 12.80 &  2.8   \\
NGC\,4775   & 12 53 45.79 & -06 37 20.1 & 1565 & SA(s)d    & 12.24 &  2.1   \\
NGC\,4904   & 13 00 56.97 & -00 01 31.9 & 1204 & SB(s)cd   & 12.60 &  2.2   \\
NGC\,5068   & 13 18 54.60 & -21 02 19.7 &  607 & SB(s)d    & 10.52 &  7.2   \\
NGC\,5585   & 14 19 48.08 & +56 43 43.8 &  571 & SAB(s)d   & 11.20 &  5.8   \\
NGC\,5584   & 14 22 23.65 & -00 23 09.2 & 1695 & SAB(rs)cd & 12.63 &  3.4  \\
NGC\,5668   & 14 33 24.30 & +04 27 02.0 & 1665 & SA(s)d    & 12.2  &  3.3  \\
\enddata
\tablecomments{Columns 1-3: object name and coordinates, as taken from
the NASA Extragalactic Database (NED). Column 4: recession velocity,
corrected according to the Virgo-centric infall model \citep{san90},
taken from the Lyon-Meudon Extragalactic Database (LEDA).
Columns 5 and 6: galaxy morphological type and apparent total B-magnitude
(NED). Column 7: galaxy major axis diameter (NED). 
}
\end{deluxetable}

\clearpage

\begin{deluxetable}{lcccccccc}
\tablewidth{0pt}
\tablecaption{Photometry of central light excess \label{tab:excess}}
\tablehead{
\colhead{(1)} & \colhead{(2)} & \colhead{(3)} & \colhead{(4)} &
\colhead{(5)} & \colhead{(6)} & \colhead{(7)} & \colhead{(8)} &
\colhead{(9)}   \\
\colhead{Galaxy} & \colhead{Distance} & \colhead{$r_{ap}$} & \colhead{Ap. Size} & 
\colhead{$\rm m_I^{total}$} & \colhead{$\rm m_I^{Exp. disk}$} &
\colhead{$\rm m_I^{Cluster}$} & \colhead{$\rm m_I^{excess}$} &
\colhead{$\rm L_I^{excess}$}\\
 & \colhead{[Mpc]} & \colhead{[arcsec]} & \colhead{[pc]} & & & & & \colhead{[$10^7\,\lsun$]}
}
\startdata
NGC\,275 & 24.0 & 2.5 & 290 & 15.33 & 15.54 & 19.47 & 17.38 & 2.9 \\
NGC\,5584 & 24.2 & 3.5 & 410 & 15.11 & 15.36 & 22.53 & 16.83 & 5.2 \\
NGC\,3346 & 18.8 & 5.0 & 460 & 14.61 & 14.73 & 19.64 & 17.21 & 2.0 \\
NGC\,2139 & 23.6 & 7.0 & 800 & 12.98 & 13.13 & 19.28 & 15.24 & 19.2 \\
\enddata
\tablecomments{Column~1: object name. Column~2: distance, derived
from the recession velocity in Column~4 of Table~\ref{tab:sample} and
assuming $\rm H_o = 70\kms$. Columns~3-4:  angular and linear
radius of aperture over which the excess emission was measured.
Column~5: apparent I-band magnitude of total light within $r_{ap}$, the
aperture radius of Column~2. Column~6: estimated apparent I-band 
magnitude within $r_{ap}$ of the galaxy disk, obtained by integrating
the best-fit exponential disk model. Column~7: apparent I-band
magnitude of nuclear cluster (from paper~I). Column~8: excess emission,
as calculated by subtracting the disk and cluster contributions from the
total. Column~9: excess I-band luminosity in solar units, derived
by assuming ${\rm M}_{I,\odot} = 4.02$. 
}
\end{deluxetable}

\clearpage
\begin{figure}
\plotone{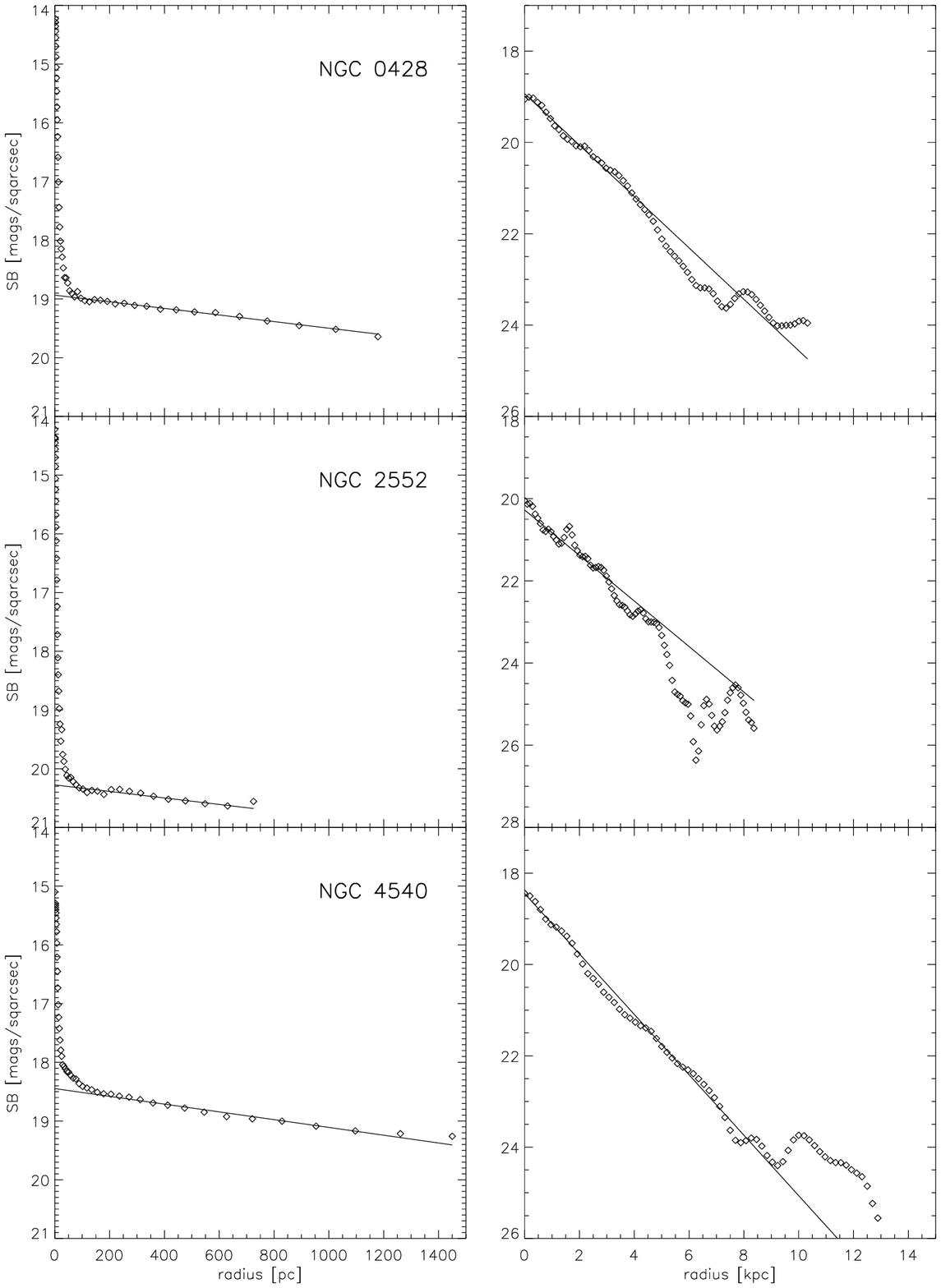}
\centerline{Fig.~\ref{fig:baghst} a)}
\end{figure}

\clearpage

\begin{figure}
\plotone{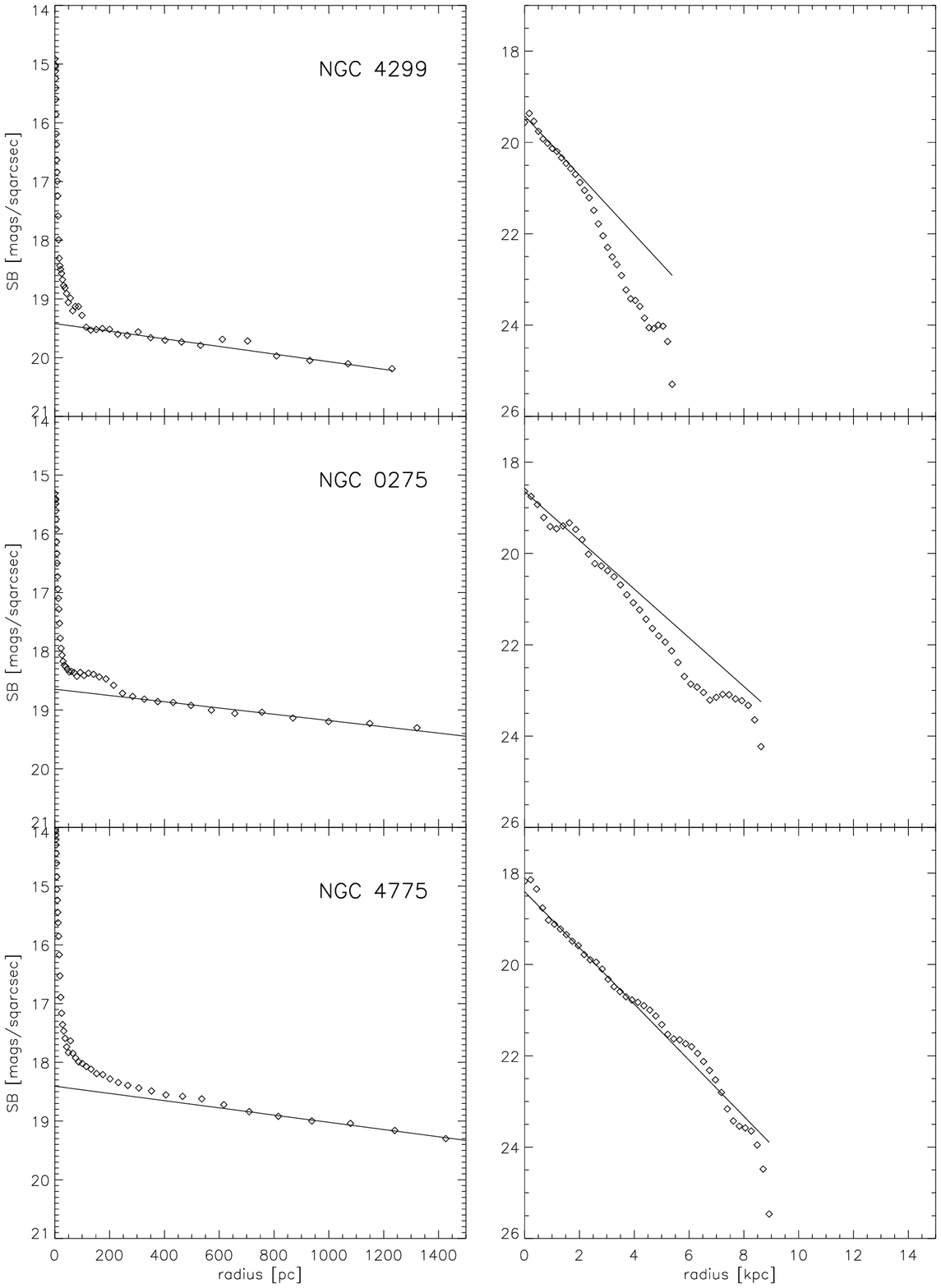}
\centerline{Fig.~\ref{fig:baghst} b)}
\end{figure}

\clearpage

\begin{figure}
\plotone{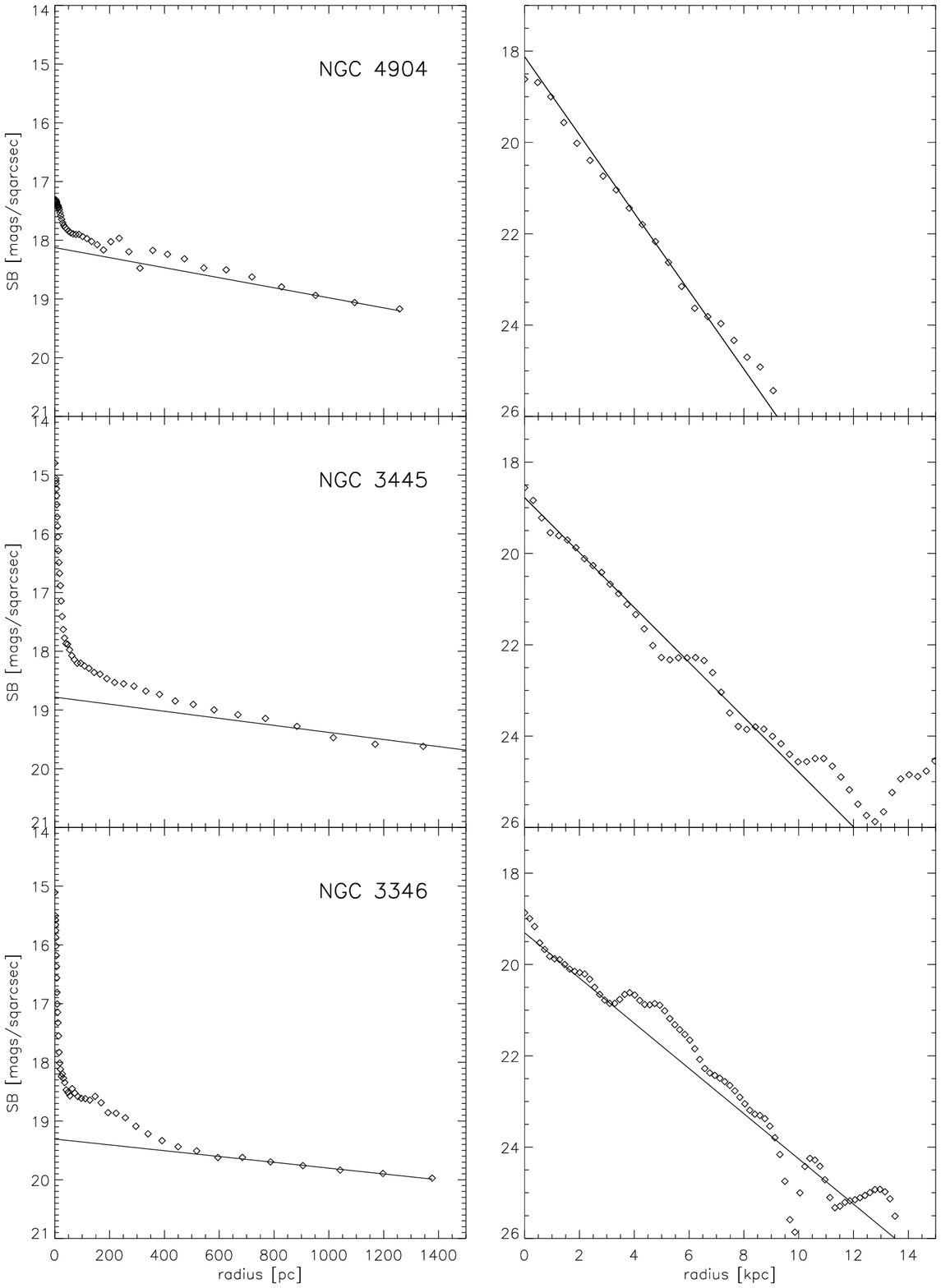}
\centerline{Fig.~\ref{fig:baghst} c)}
\end{figure}

\clearpage

\begin{figure}
\plotone{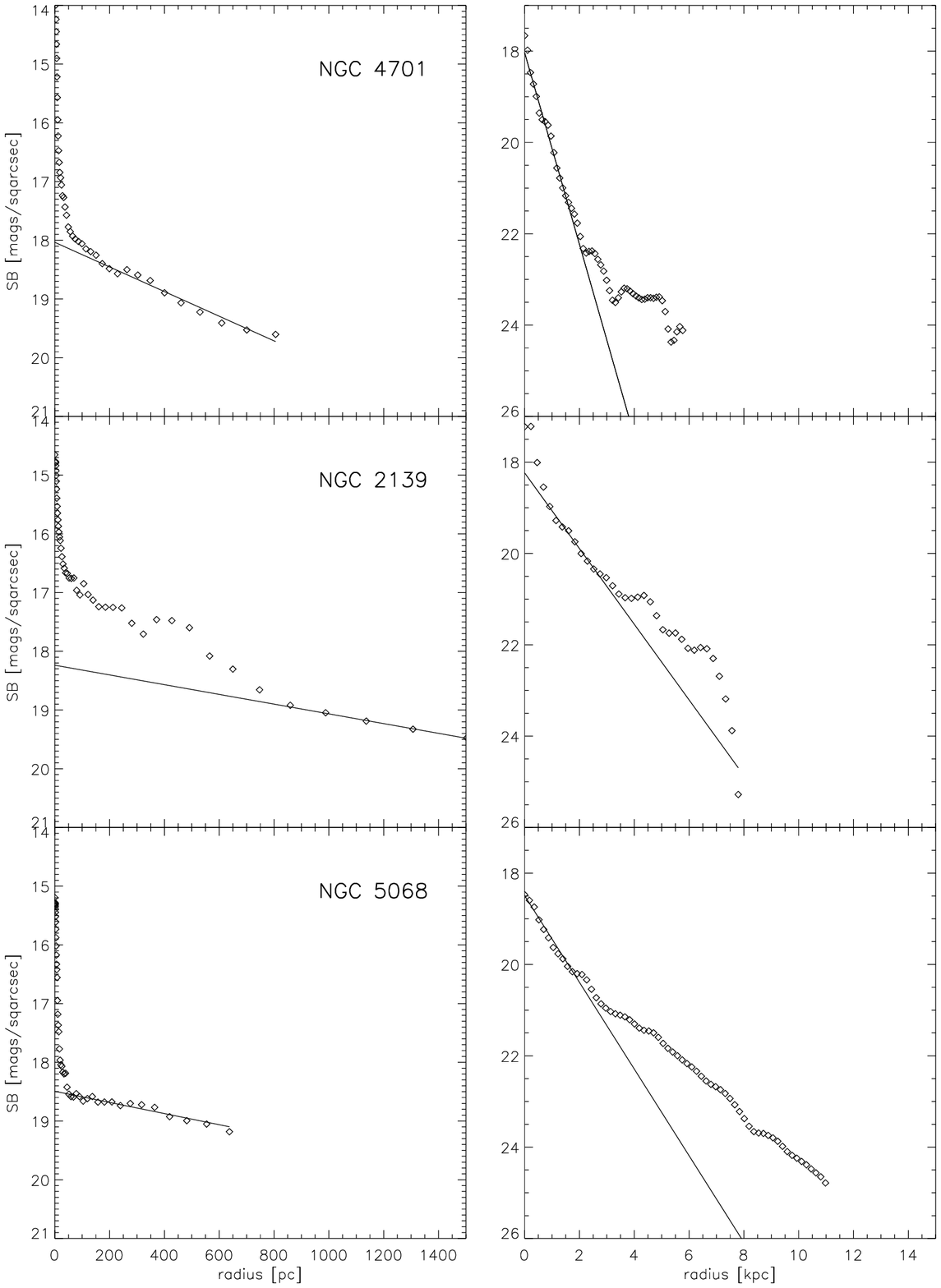}
\centerline{Fig.~\ref{fig:baghst} d)}
\end{figure}

\clearpage

\begin{figure}
\plotone{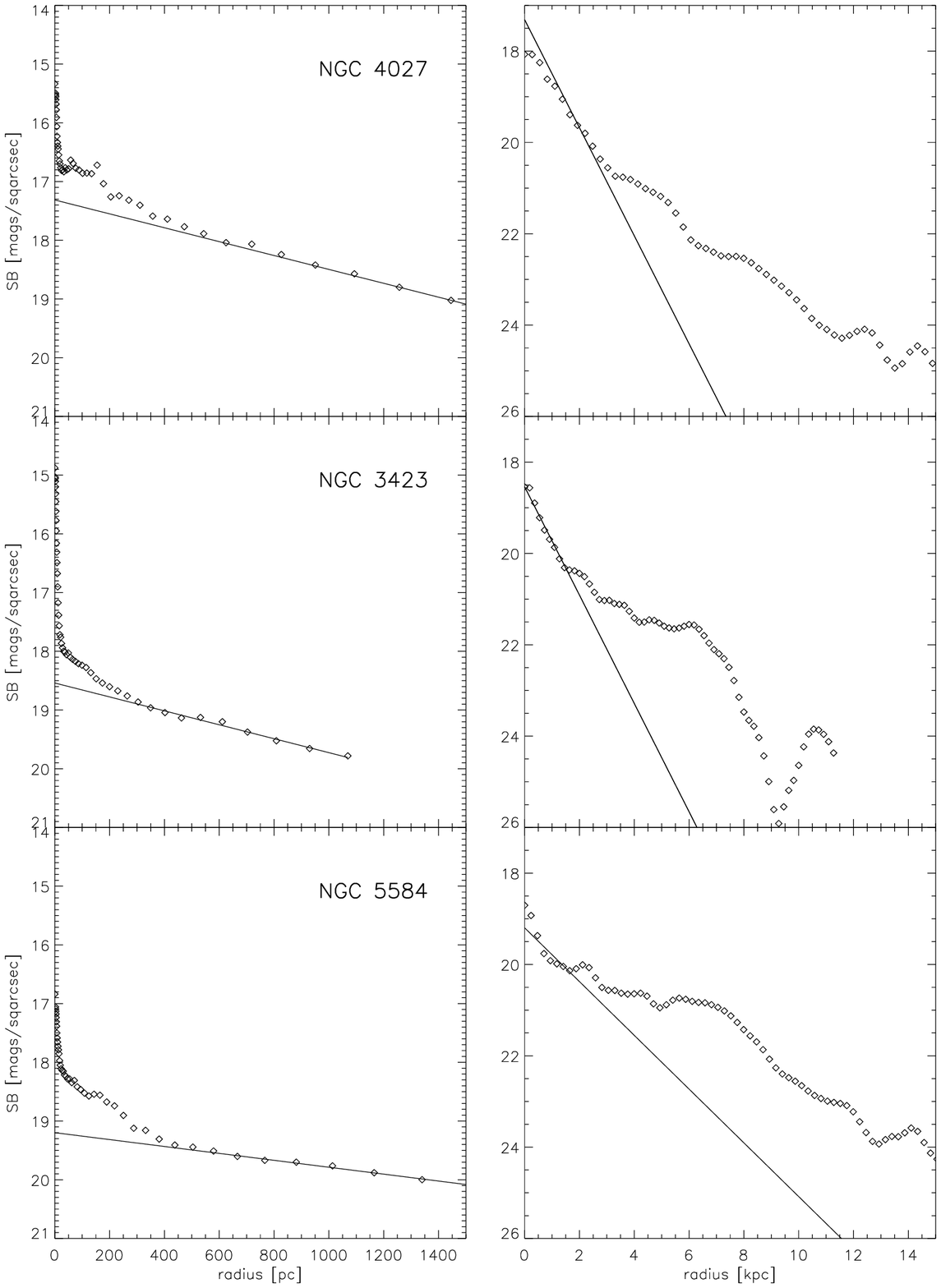}
\centerline{Fig.~\ref{fig:baghst} e)}
\end{figure}

\clearpage

\begin{figure}
\plotone{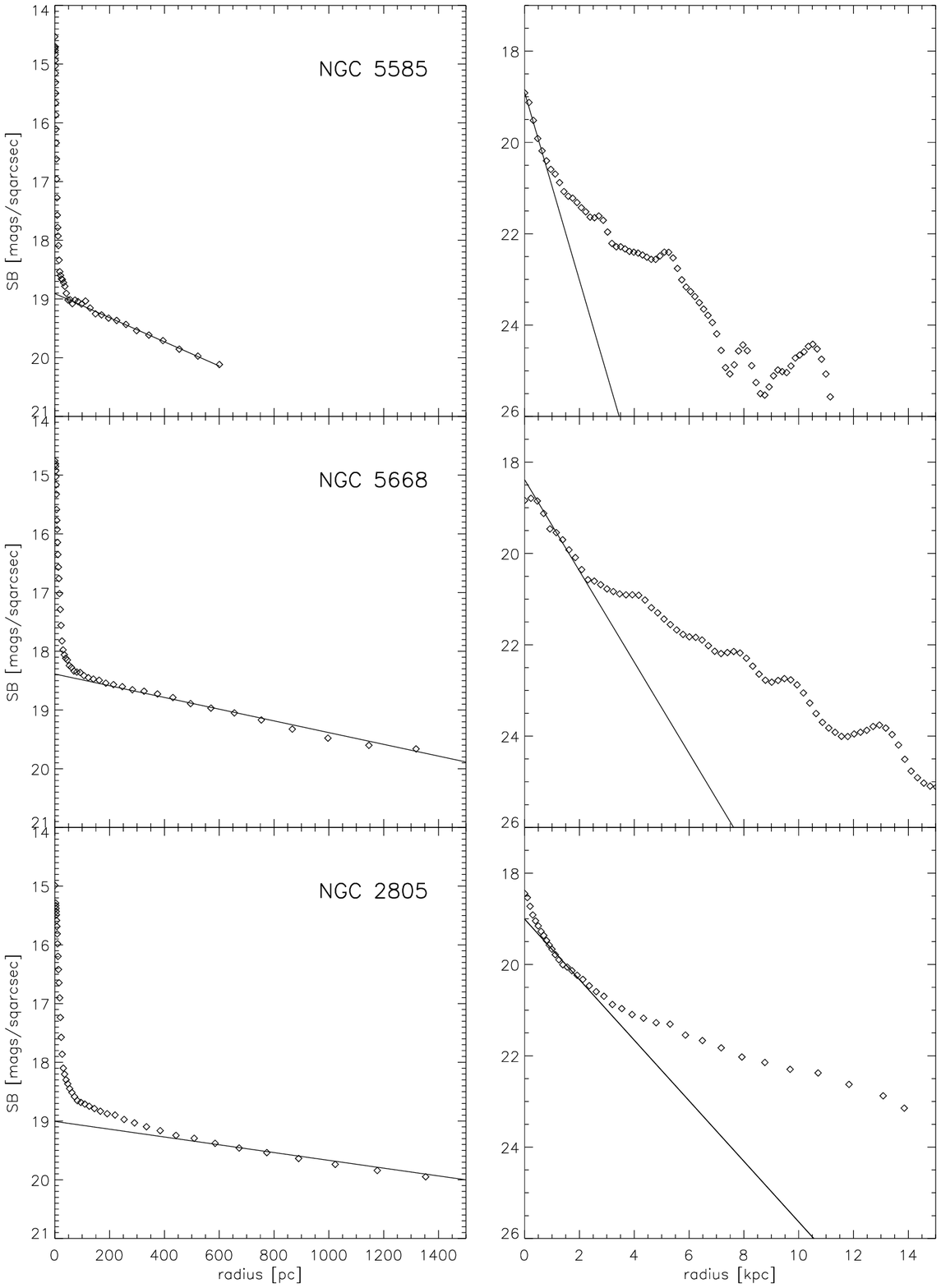}
\centerline{Fig.~\ref{fig:baghst} f)}
\end{figure}

\clearpage

\begin{figure}
\plotone{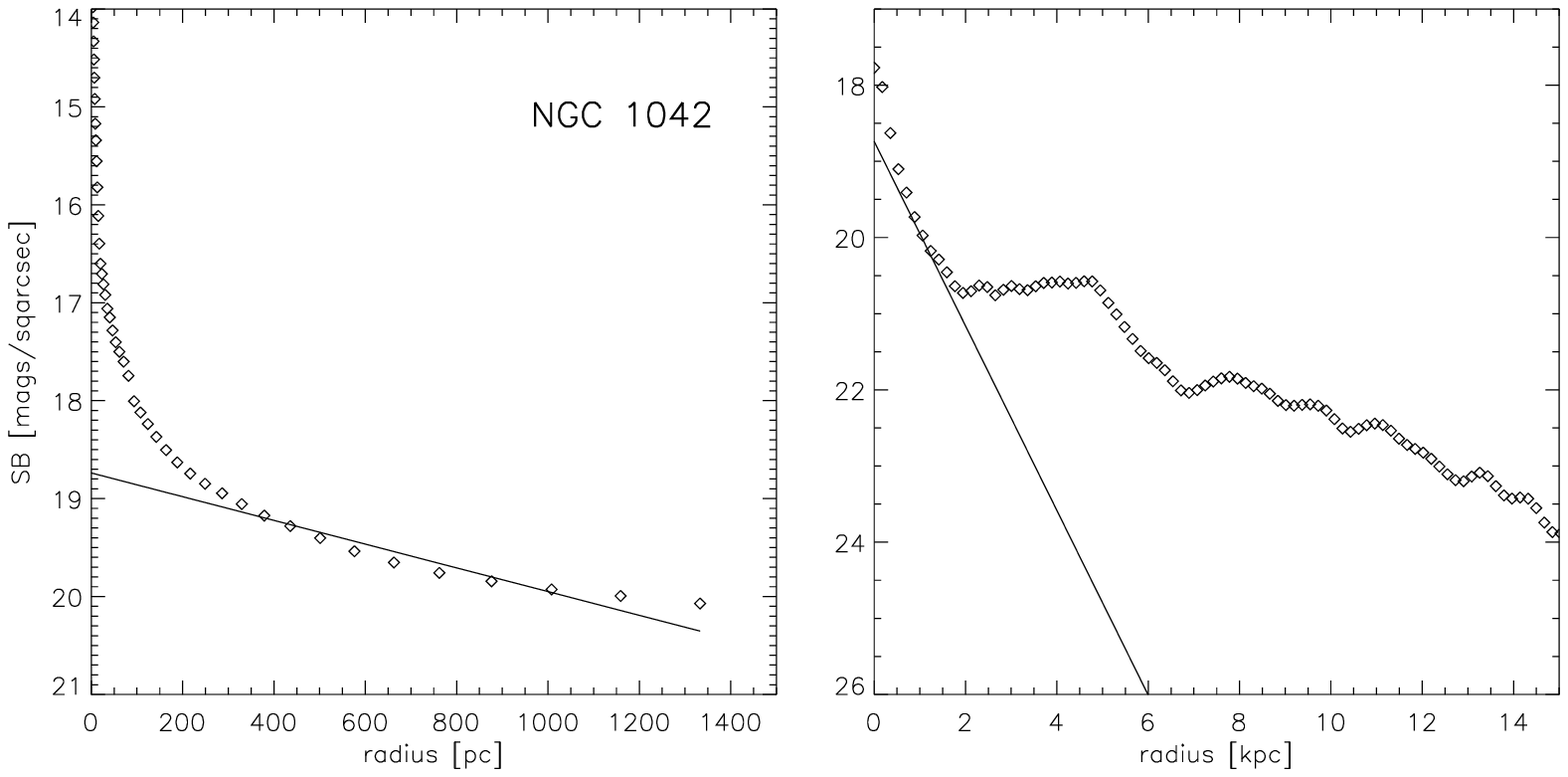}
\centerline{Fig.~\ref{fig:baghst} g)}
\end{figure}

\clearpage

\clearpage
\begin{figure}
\plottwo{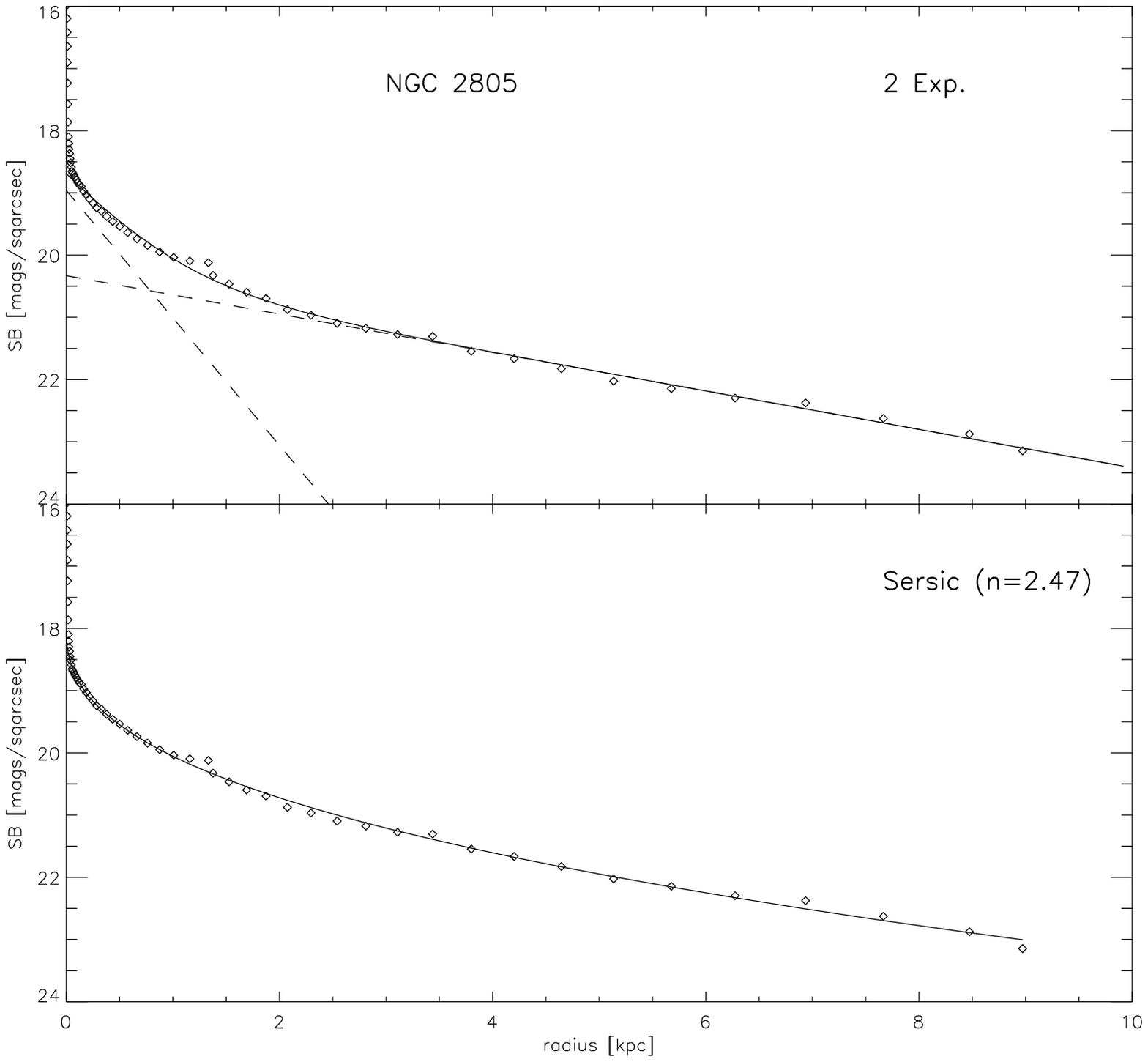}{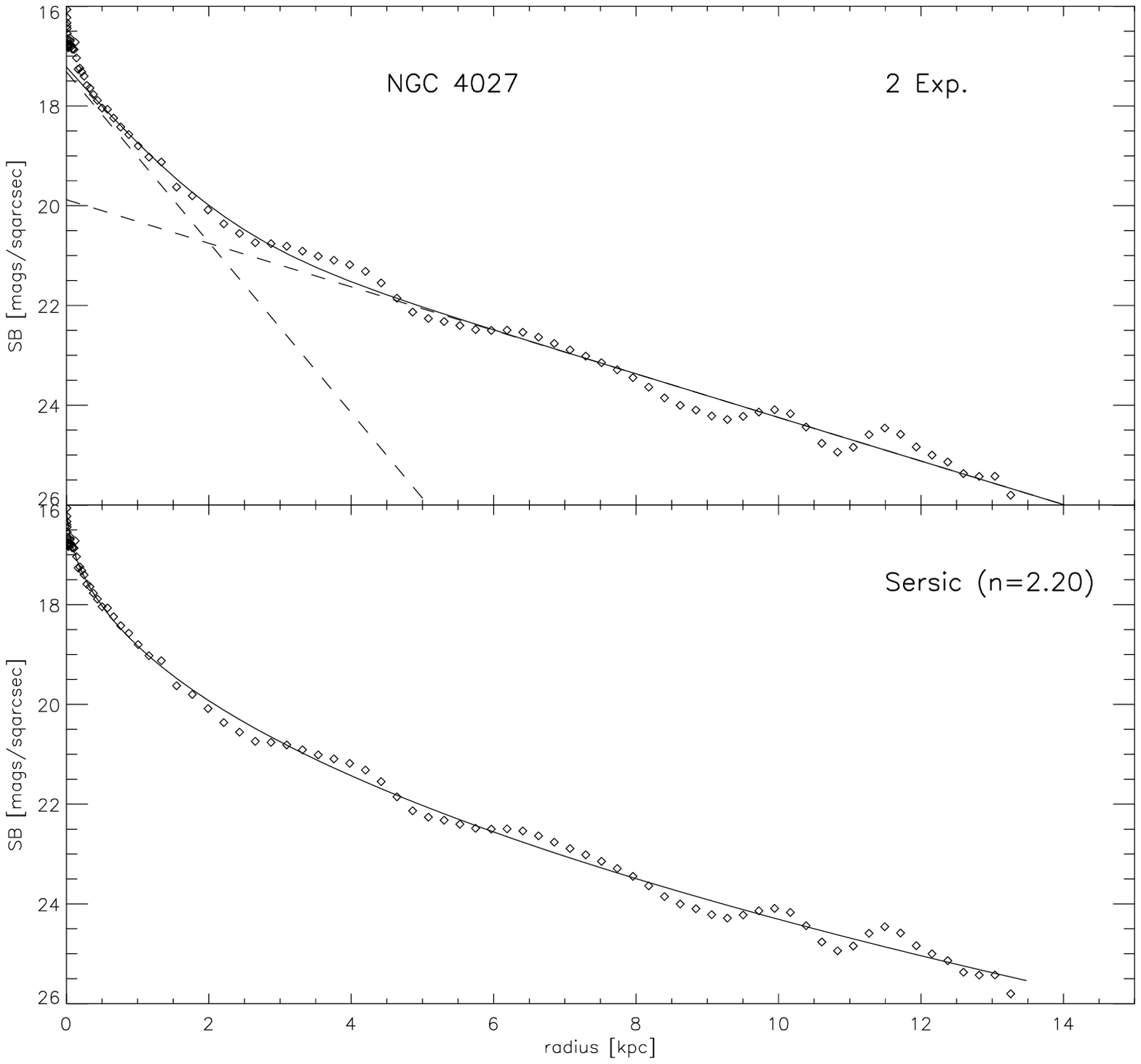}
\end{figure}

\begin{figure}
\plottwo{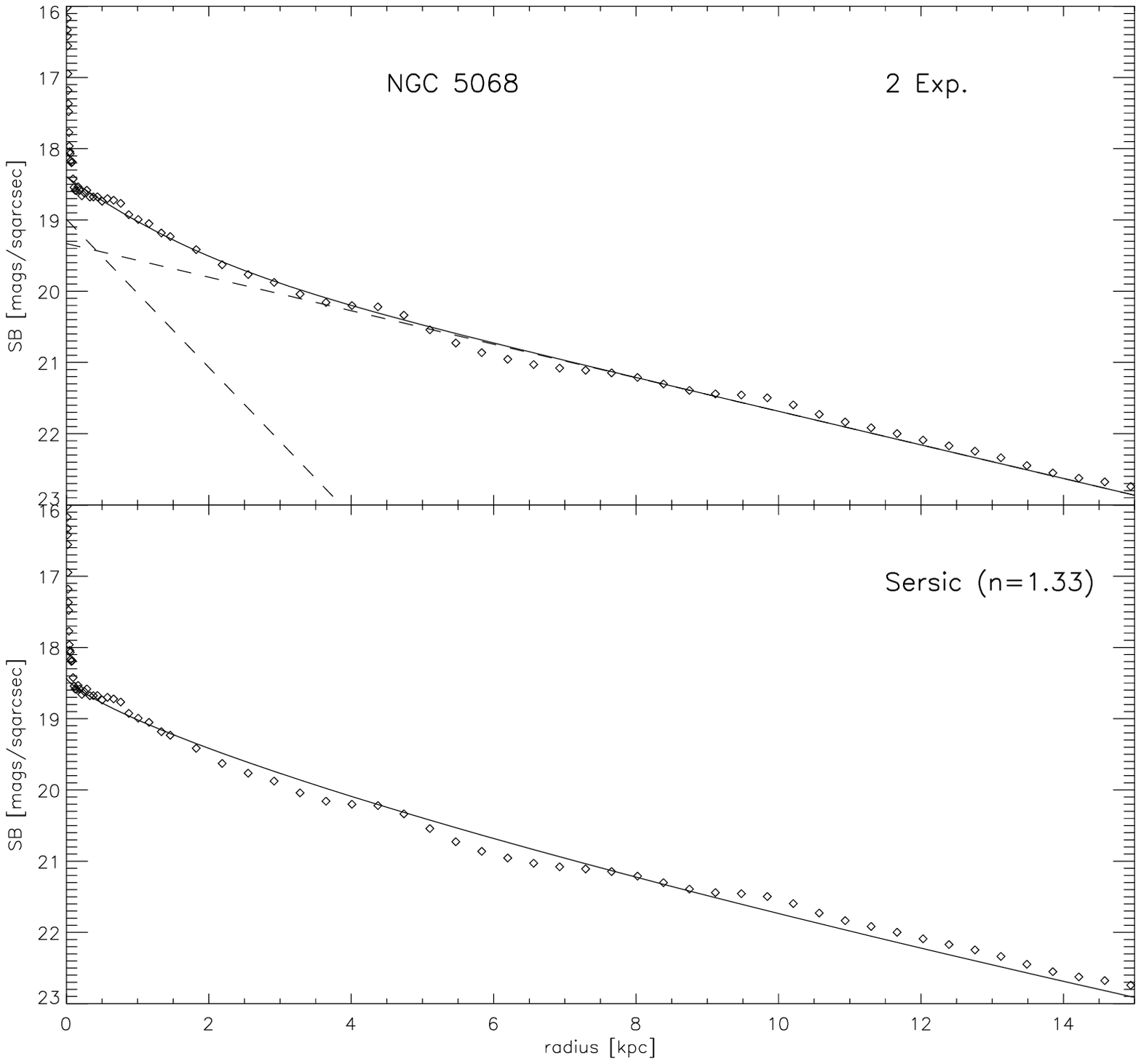}{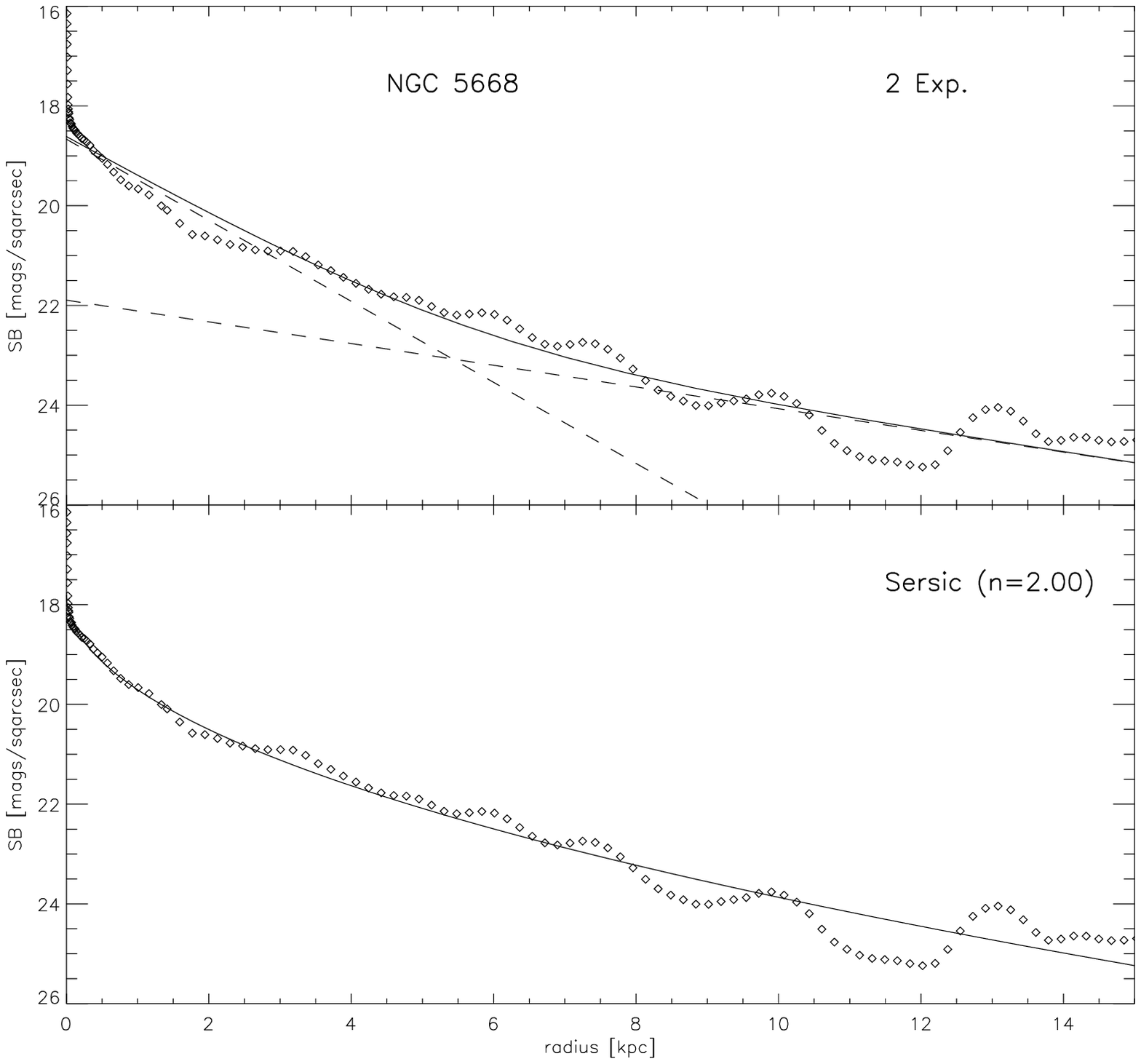}
\centerline{Fig.~\ref{fig:comps}}
\end{figure}

\begin{figure}
\plotone{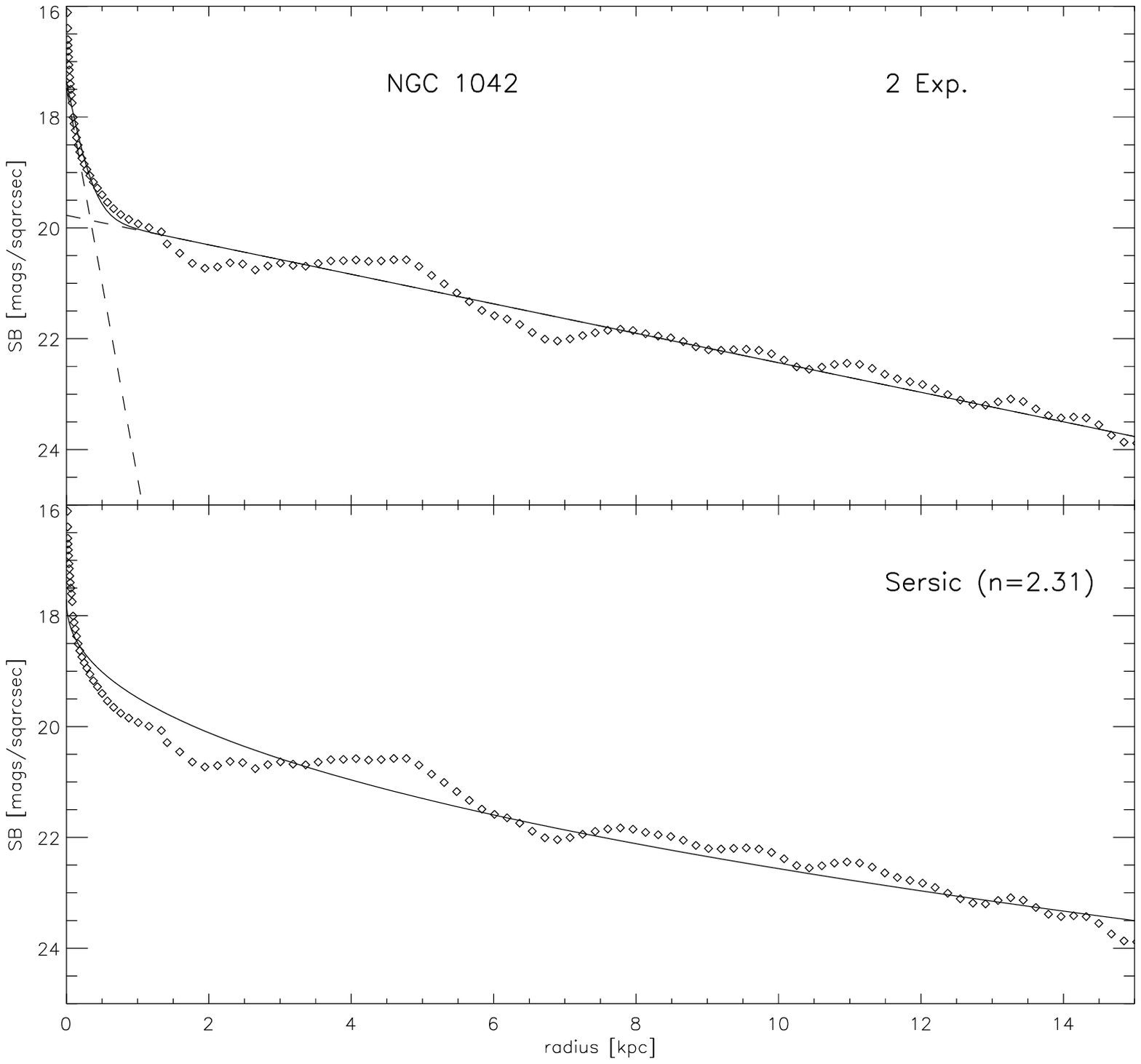}
\centerline{Fig.~\ref{fig:comps}- cont.}
\end{figure}


\end{document}